\documentclass[aps,pra,twocolumn,groupedaddress,amsmath,amssymb,superscriptaddress]{revtex4-1}
\usepackage{graphicx}  
\usepackage{dcolumn}   
\usepackage{bm}        
\usepackage{verbatim}   
\usepackage{ mathrsfs }
\usepackage[colorlinks=true,linkcolor=blue,citecolor=blue,allcolors=blue]{hyperref}
\usepackage{float}
\numberwithin{equation}{section}

\begin{document}

\title{Tunneling of two interacting atoms from excited states}

\author{I.S. Ishmukhamedov}
\email{i.ishmukhamedov@mail.ru}
\affiliation{Bogoliubov Laboratory of Theoretical Physics, Joint Institute for Nuclear Research, Dubna, Moscow Region 141980, Russian Federation}
\affiliation{Al-Farabi Kazakh National University, Almaty 050040, Kazakhstan}
\affiliation{Institute of Nuclear Physics, the Ministry of Energy of the Republic of Kazakhstan, Almaty 050032, Republic of Kazakhstan}

\author{A.S. Ishmukhamedov}
\email{ishmukhamedov.altay@gmail.com}
\affiliation{Al-Farabi Kazakh National University, Almaty 050040, Kazakhstan}
\affiliation{Institute of Hydrogeology and Geoecology, the Ministry of Education and Science, Almaty 050010, Republic of Kazakhstan}

\begin{abstract}
  We consider tunneling of two interacting atoms with an even spatial symmetry. The atoms are prepared in two lowest excited states with respect to relative and center-of-mass motions. We observe monotonic and non-monotonic dependence of the decay rate of the total probability as a function of the interatomic coupling strength $g$. We find a transition from uncorrelated to correlated pair tunneling as a function of $g$ and an external trap barrier. The similar system has been investigated for two interacting $^6$Li atoms in the deterministic Heidelberg experiment.
\end{abstract}

\maketitle
\section{INTRODUCTION}

Understanding of quantum tunneling is a key question which arise in various fields of physics such as a decay of an $\alpha$ particle \cite{migdal,maruyama,scamps} from a nucleus, superfluidity of electrons in metals \cite{gennes} and $^3$He atoms \cite{leggett}. It is still, however, challenging to include a correlation into an analysis of few-body systems \cite{rontani,rontani2,gharashi,maruyama}.

Only recently it has become possible for the experimental realization of a pure two-body quantum system \cite{serwane,zuern,zuern2}. This has been done in the setup of the Heidelberg group in which they prepared two cold $^6$Li atoms in the ground and lowest excited states of a dipole trap. Due to tight confinement in the transverse direction this system can be considered as a one-dimensional one \cite{idziaszek}. By imposing an additional linear magnetic field, the atoms start to escape from the trap due to quantum tunneling mechanism through the trap barrier. The interatomic coupling strength is varied by means of a magnetic Feshbach resonance \cite{chin} in which the interaction could be varied in a wide range and could be attractive or repulsive.

There were several theoretical methods to describe the tunneling dynamics of this system \cite{rontani,rontani2, gharashi}. However, they mostly considered the tunneling from the ground state and only from few levels of the lowest excited states. In the present paper we extend these analyses and add more excited states, which also consist of a center-of-mass excited state branch. We should also note similar works on quantum tunneling \cite{nesterenko, krass, lode, hunn, dobr}

The paper is organized as follows. Sec.\ref{model} describes the model Hamiltonian and a numerical method to solve the time-dependent Schr\"{o}dinger equation. Sec.\ref{results} discusses the spectrum for the initial state and tunneling dynamics of upper and lower excited state branches. Sec.\ref{conclusion} summarizes the results.

\section{MODEL AND THE NUMERICAL METHOD}
\label{model}
Tunneling dynamics of the two-atom system is modeled by the time-dependent Schr\"{o}dinger equation
\begin{eqnarray}\label{tdse}
i\hbar\frac{\partial \psi(x_1,x_2,t)}{\partial t}=H(x_1,x_2)\psi(x_1,x_2,t),
\end{eqnarray}
where the two-body Hamiltonian reads
\begin{eqnarray}\label{ham}
H(x_1,x_2)=H_1(x_1)+H_2(x_2)+V^{(\textsf{aa})}(x_1-x_2).
\end{eqnarray}
Here
\begin{eqnarray}\label{sah}
H_j(x_j)=-\frac{\hbar^2}{2m}\frac{\partial^2}{\partial x_j^2}+V^{(\textsf{at})}(x_j),\qquad j=1,2.
\end{eqnarray}
are the single-particle Hamiltonians, in which $V^{(\textsf{at})}$ describe the atom-trap interaction and $m$ is the atomic mass.

\begin{figure}
\centering
\includegraphics[width=6cm,clip]{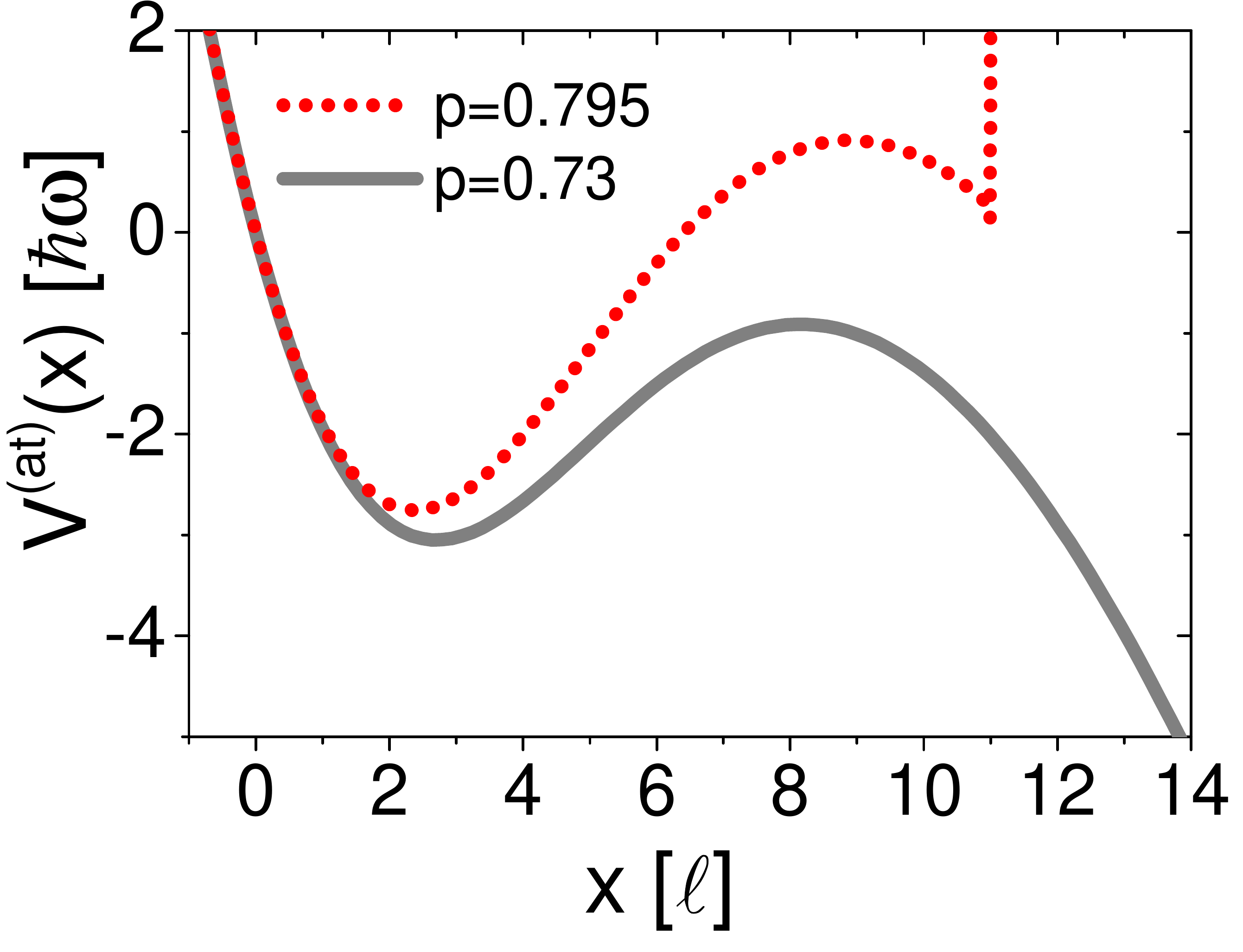}
\caption{Plot of the trapping potential $V^{(\textsf{at})}(x)$: $p=0.795$ (dashed line) plus a hard wall at $x=11\ell$ [represents the trap at $t=0$] and $p=0.73$ (solid line) [represents the trap at $t>0$]}\label{trap_fig}
\end{figure}

The atom-trap potential is taken from \cite{gharashi,zuern,zuern2} which represents an optical trap and magnetic field gradient (Fig.\ref{trap_fig}):
\begin{equation}\label{trap}
V^{(\textsf{at})}(x)=pV_0\left[ 1-\frac{1}{\left(x/x_R\right)^2+1} \right]-\mu_B \mathcal{C}x,
\end{equation}
where $\mu_B$ is the Bohr magneton, $x_R=8.548\ell~(\ell=\sqrt{\hbar/m\omega})$ is the Rayleigh range, $V_0=56.16\hbar\omega$ is the maximum depth of the optical trap, $p$ is the optical trap depth, $\mathcal{C}$ depends on external magnetic-field strength, magnetic-field gradient and the hyperfine state of the atoms, $\omega=2\pi\times1234$ Hz is the trap frequency. These values represents realistic trap parameters used in \cite{zuern,zuern2,gharashi}. We take $p=0.795$ for the initial state which corresponds to the actual value used in the experiment \cite{zuern,zuern2,gharashi}. At $t>0$ we reduce the barrier width by taking $p=0.73$ (if not stated otherwise). This value ensures a fast enough tunneling time scale from the trap potential. The parameter $\mathcal{C}$ will be defined in the following sections.

For the interatomic interaction potential $V^{(\textsf{aa})}(z_1-z_2)$ we choose the Gaussian shape \cite{gharashi, ishmukh2}
\begin{eqnarray}\label{aa}
V^{(\textsf{aa})}(x_1-x_2)=-V_G\exp\left\{-\frac{(x_1-x_2)^2}{2r_0^2}\right\},
\end{eqnarray}
where $V_G$ and $r_0$ - are the depth and range of the interatomic interaction \eqref{aa}. The value $r_0=0.1\ell$ \cite{ishmukh2, gharashi} - well represents the short-range interaction, which atoms experience at low energies. To parameterize the interaction \eqref{aa} we use the 1D contact coupling strength $g$ which has a simple relation with a 1D scattering length $a$ as $g=-2\hbar^2/(ma)$. To compute $a$ we solve the scattering problem for the relative motion of the two-atom system in the absence of the trapping potential \cite{ishmukh, ishmukh2}.

In order to integrate \eqref{tdse} we use the split-operator method, which is based on ideas \cite{marchuk} and has been developed in the works \cite{melezhik, melezhik2,melezhik3} in application to confined ultracold atom-atom collisions in waveguide-like traps:
\begin{eqnarray}\label{split}
\nonumber
&&\psi(x_1,x_2,t+\Delta t)=\exp\left\{-i \frac{\Delta t}{2\hbar} V^{(\textsf{aa})}(x_1-x_2)\right\}
\\ \nonumber
&&\exp\left\{-\frac{i \Delta t H_1(x_1)}{\hbar}\right\}\exp\left\{-\frac{i \Delta t H_2(x_2)}{\hbar}\right\}
\\
&&\exp\left\{-i \frac{\Delta t}{2\hbar} V^{(\textsf{aa})}(x_1-x_2)\right\}
\psi(x_1,x_2,t)
\end{eqnarray}
The action of the operators $\exp\left\{-i \Delta t H_j(x_j)\right\}$ is approximated by the Crank-Nicolson scheme, which maintains the accuracy order of the split-operator scheme \eqref{split} to $\mathcal{O}(\Delta t^3)$. The partial derivatives in \eqref{sah} are approximated by the sixth-order finite-differences.

At large $x_1$ and $x_2$ the wave function should has a shape of an outgoing wave. This boundary condition can be modeled by introducing into the original Hamiltonian \eqref{ham} a complex absorbing potential (CAP) $iW(x_j)$ \cite{meyer,lode}, which absorbs the wave-packet at the edges of the simulation grid, of the form:
\begin{eqnarray}\label{cap}
W(x_j)=w_c(|x_j|-x_c)^2\theta(|x_j|-x_c),\qquad \hspace{-.5cm}j=1,2,
\end{eqnarray}
where $\theta(x)$ is the Heaviside step function. Final results should not depend on a particular choice of the parameters $w_c$ and $x_c$. Therefore, there is no unique choice of these parameters, but rather a domain where the final results do not significantly change under the variation of them. We find that $w_c=-1\hbar\omega\ell^{-2}$ and $x_c=25\ell$ lie in that domain.

\section{RESULTS}
\label{results}

\subsection{Initial solution and the energy spectum}
\begin{figure}[H]
\centering
\includegraphics[width=6cm,clip]{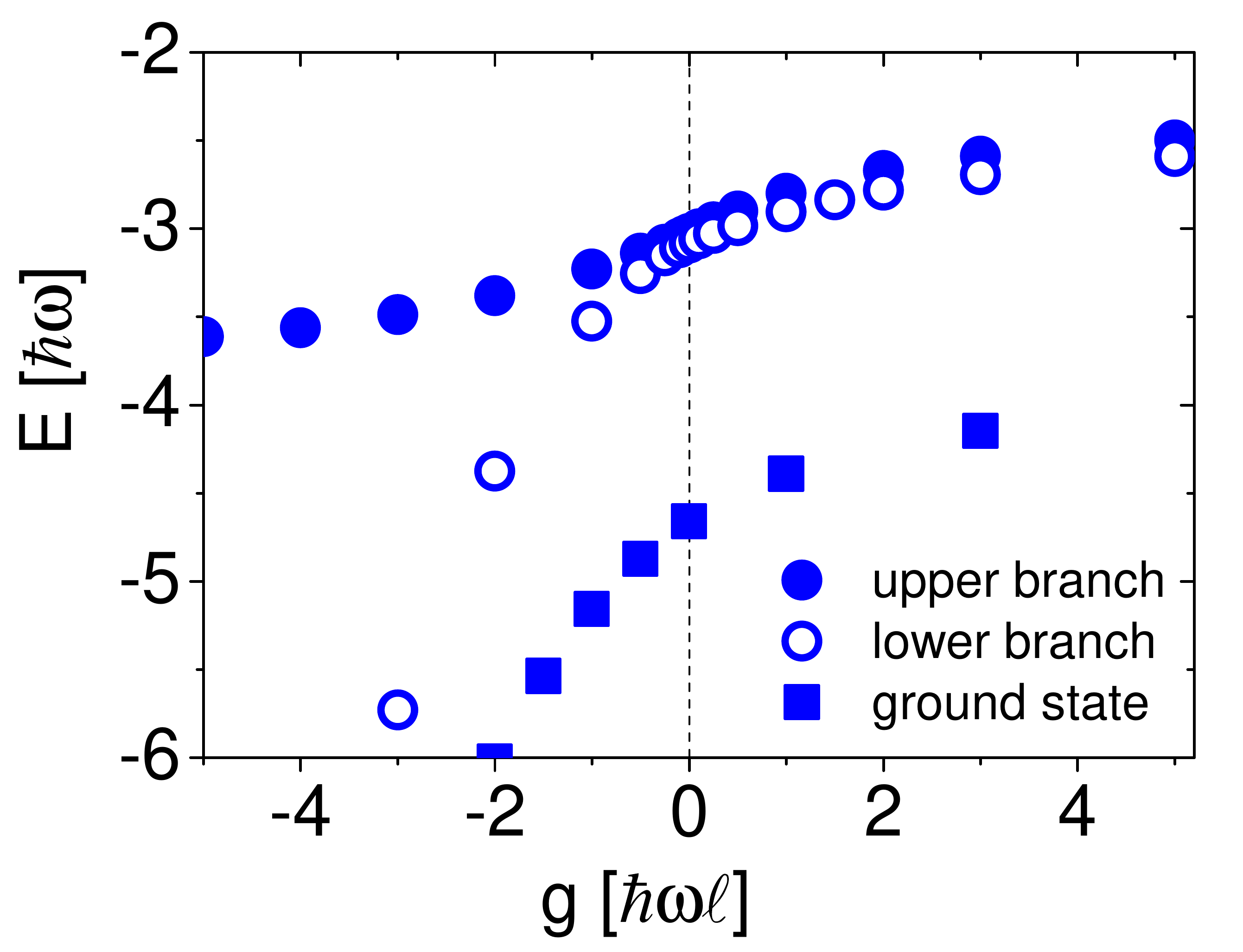}
\caption{Energy spectrum of the "closed" trapping potential \eqref{trap} as a function of the coupling strength $g$, using $p=0.795$ and putting a hard wall at $x=11\ell$. Upper and lower excited state branches correspond to the doubly excited states with respect to relative and center-of-mass motions \cite{ishmukh}.}\label{energy}
\end{figure}

\begin{figure}[H]
\centering
\begin{tabular}{cccc}
g=-1 & g=0 & g=1 & g=3\\
\multicolumn{4}{c}{Upper branch}\\
\includegraphics[width=.25\textwidth]{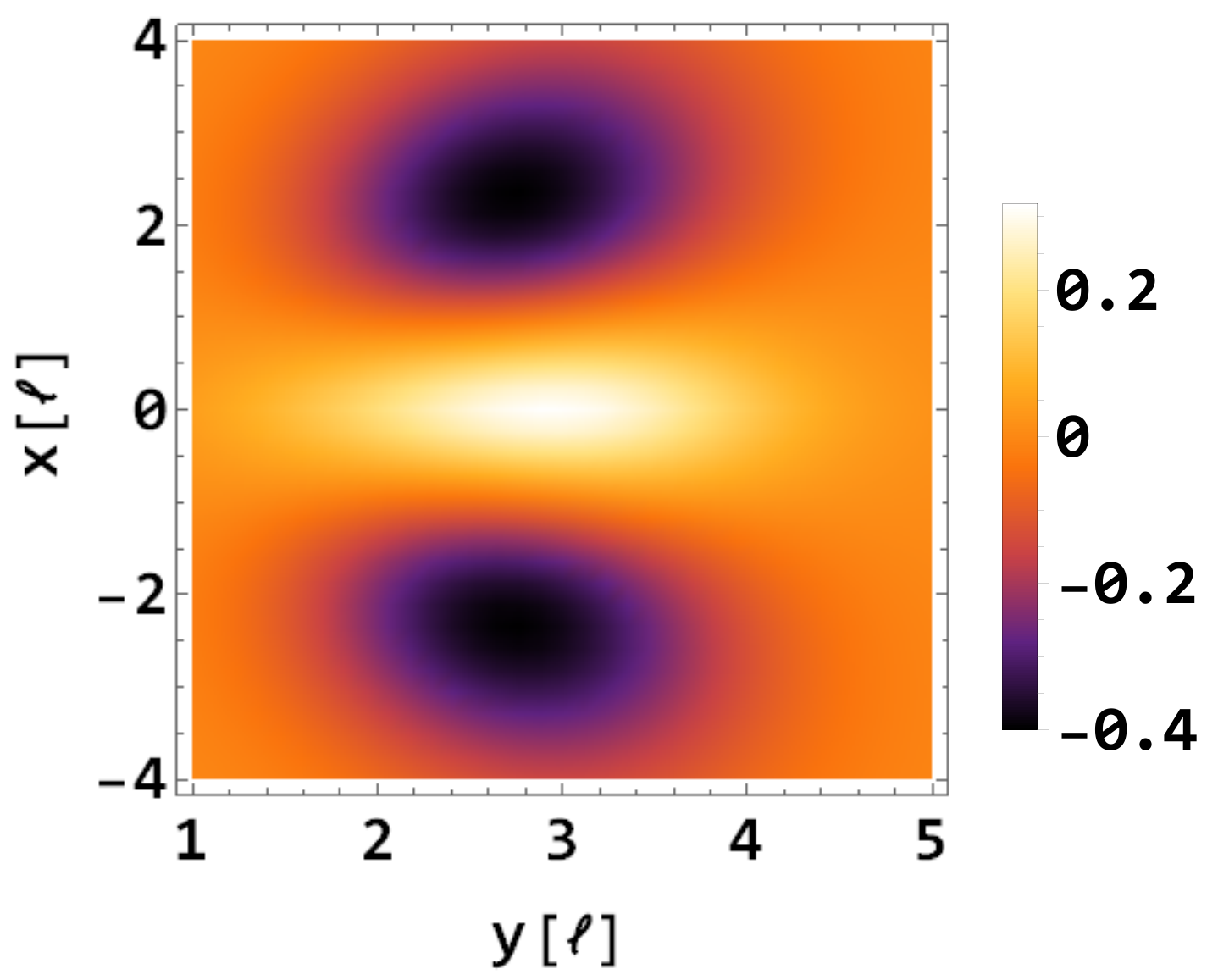} &
\includegraphics[width=.25\textwidth]{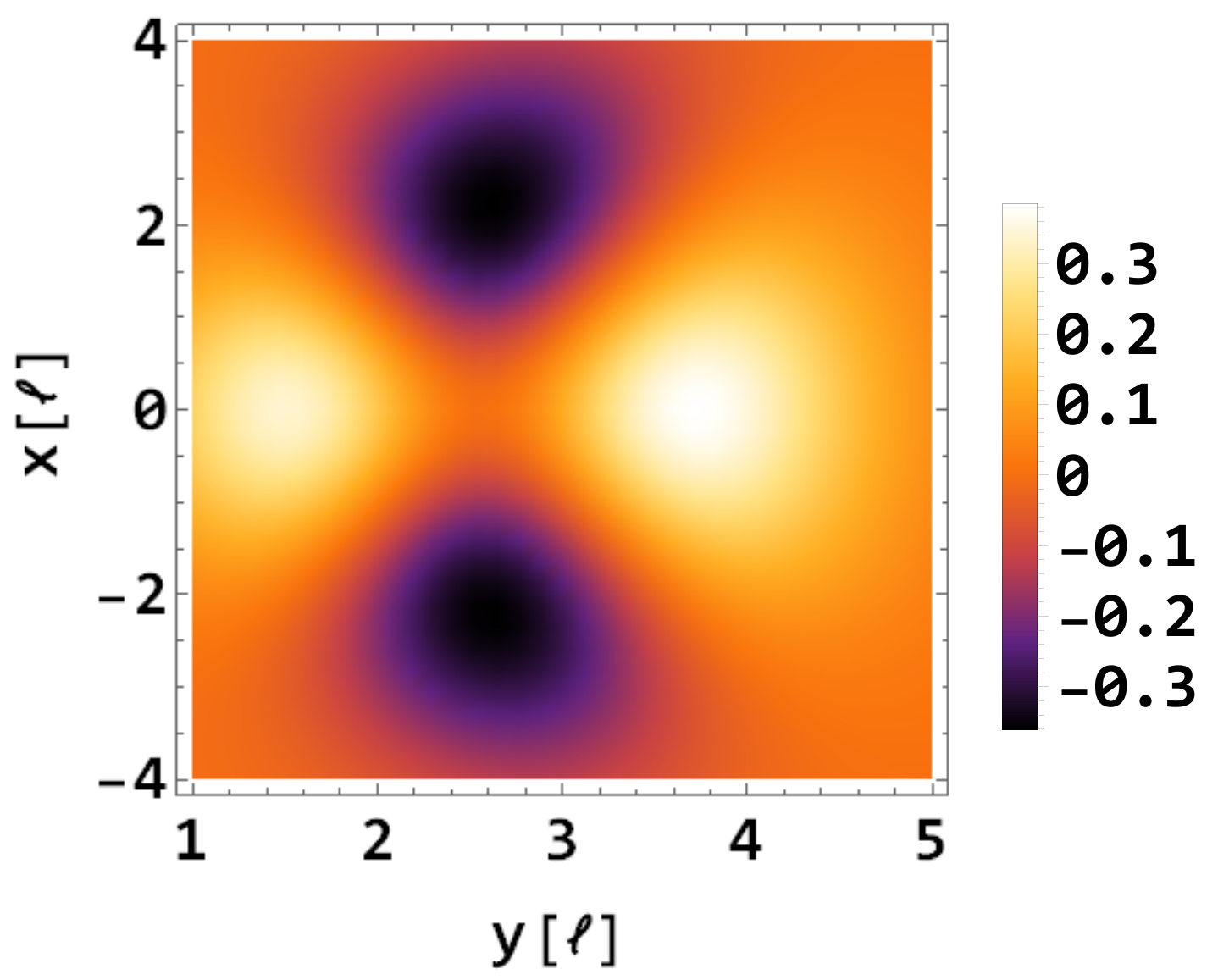} &
\includegraphics[width=.25\textwidth]{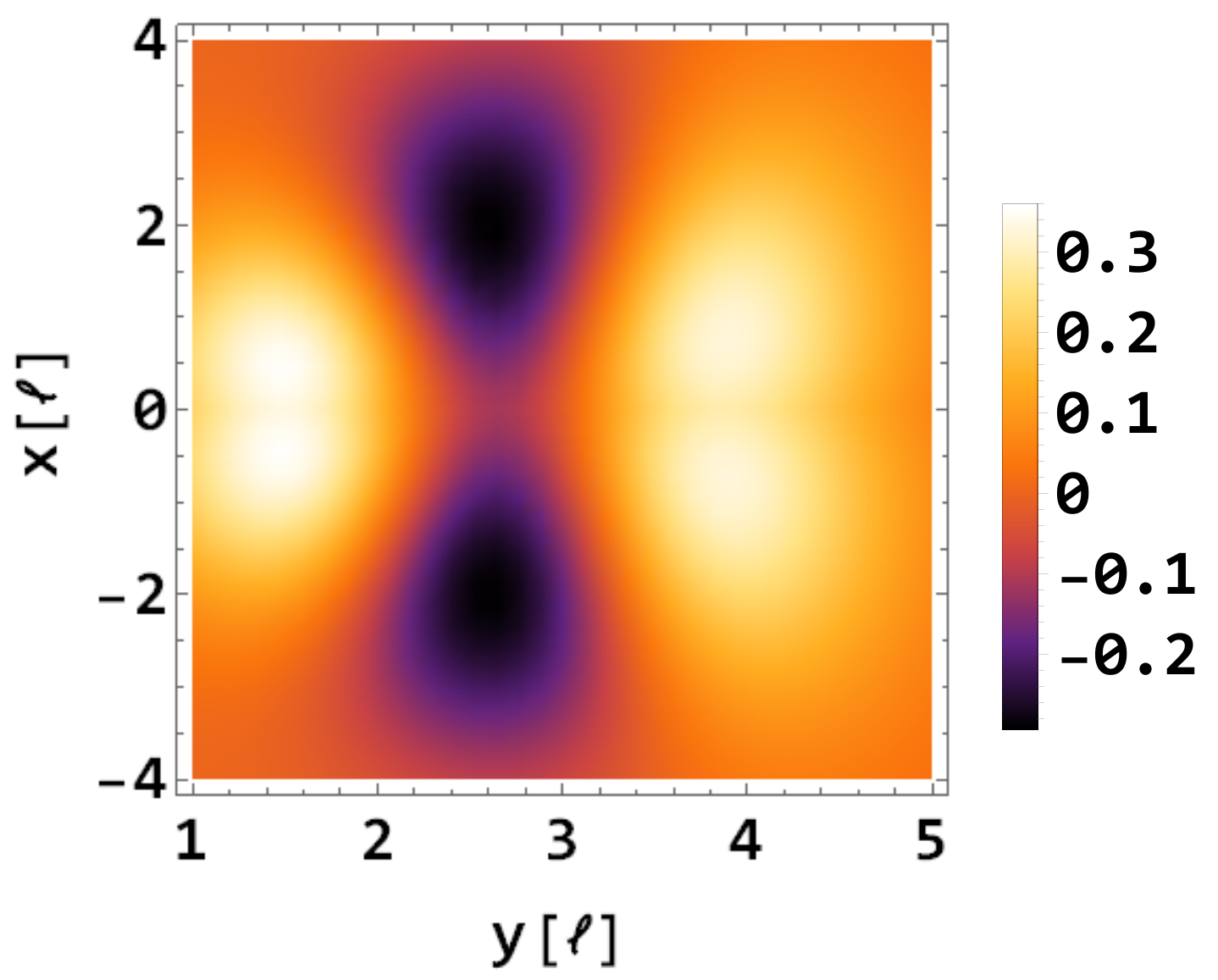} &
\includegraphics[width=.25\textwidth]{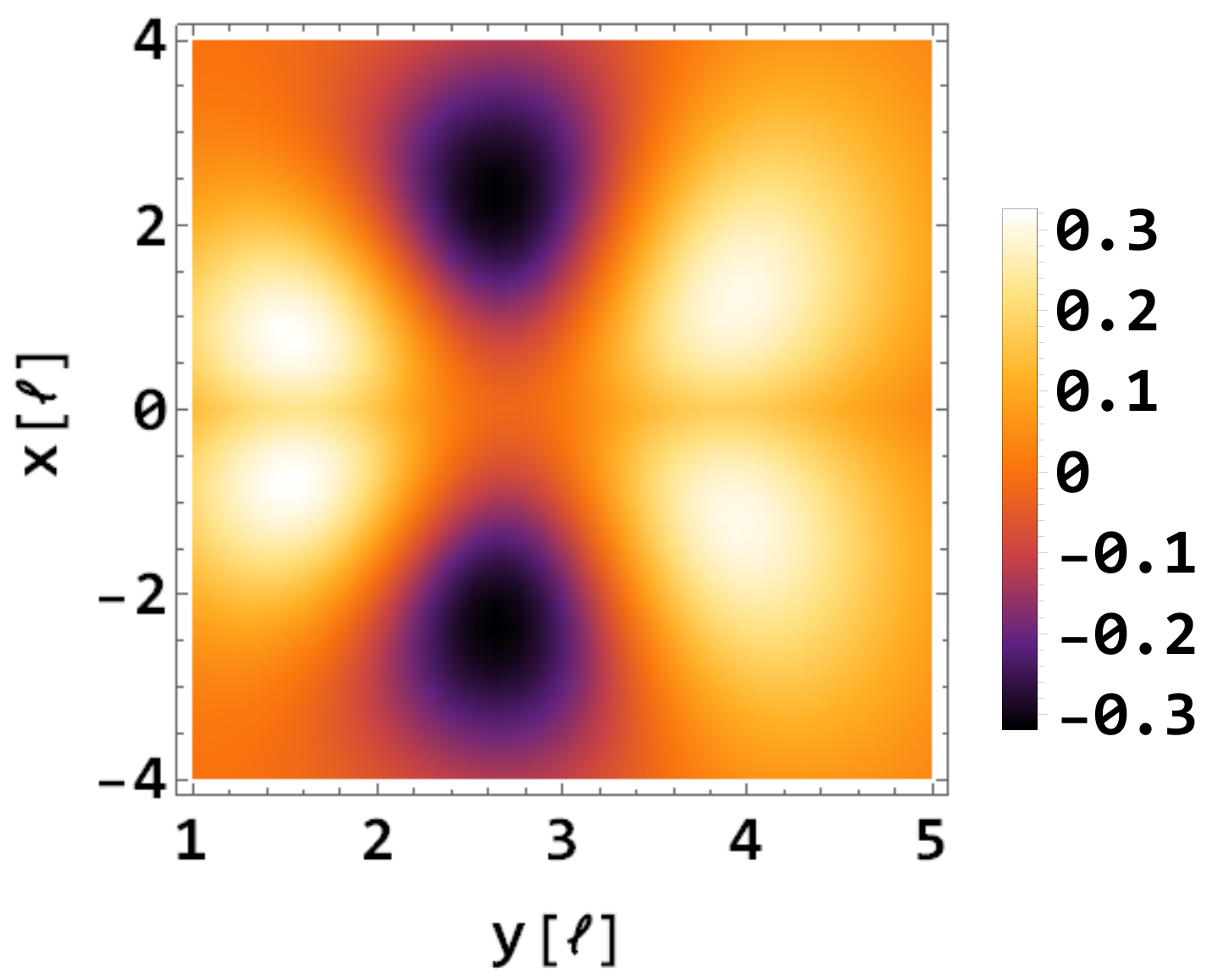} \\
\multicolumn{4}{c}{Lower branch}\\
\includegraphics[width=.25\textwidth]{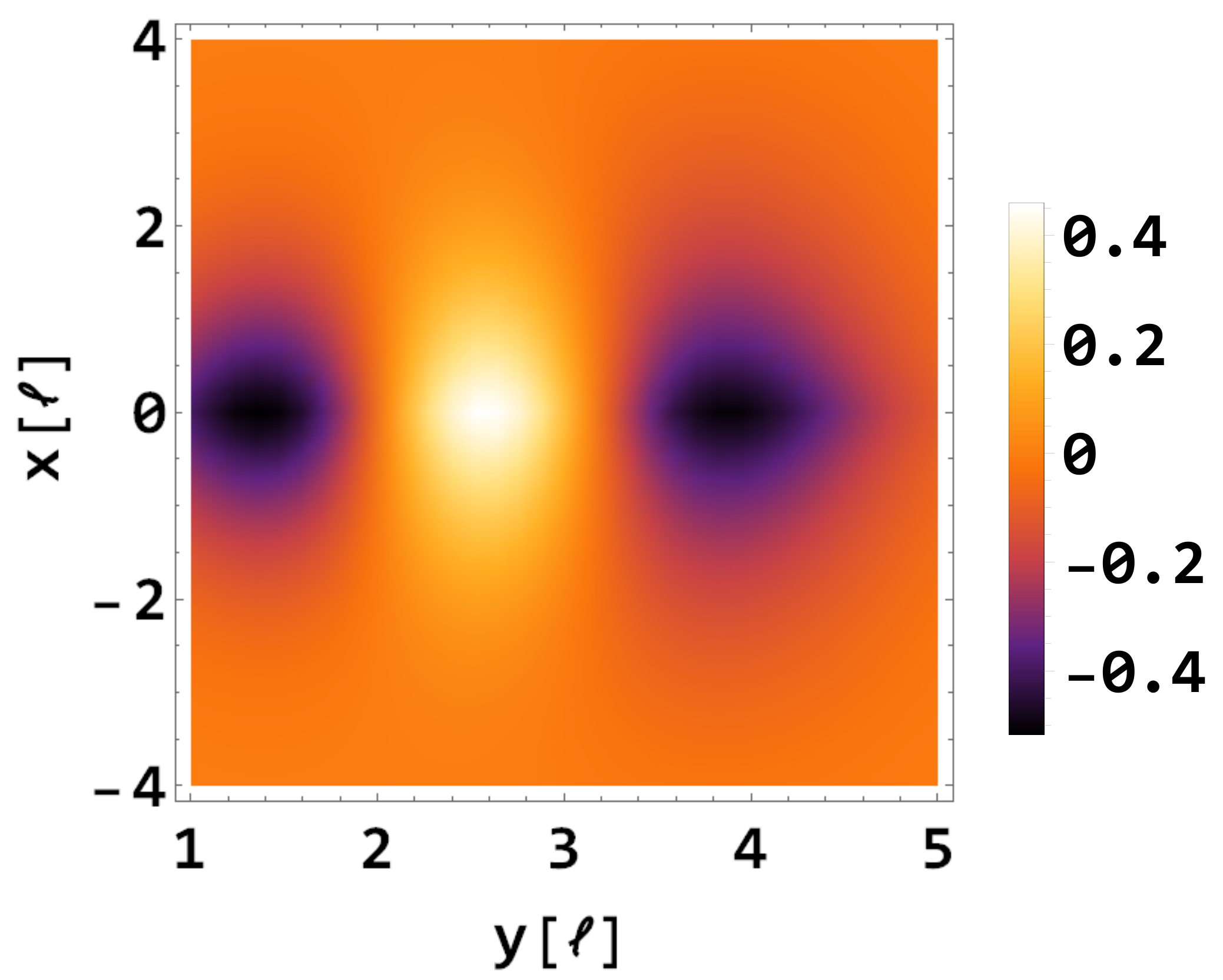} &
\includegraphics[width=.25\textwidth]{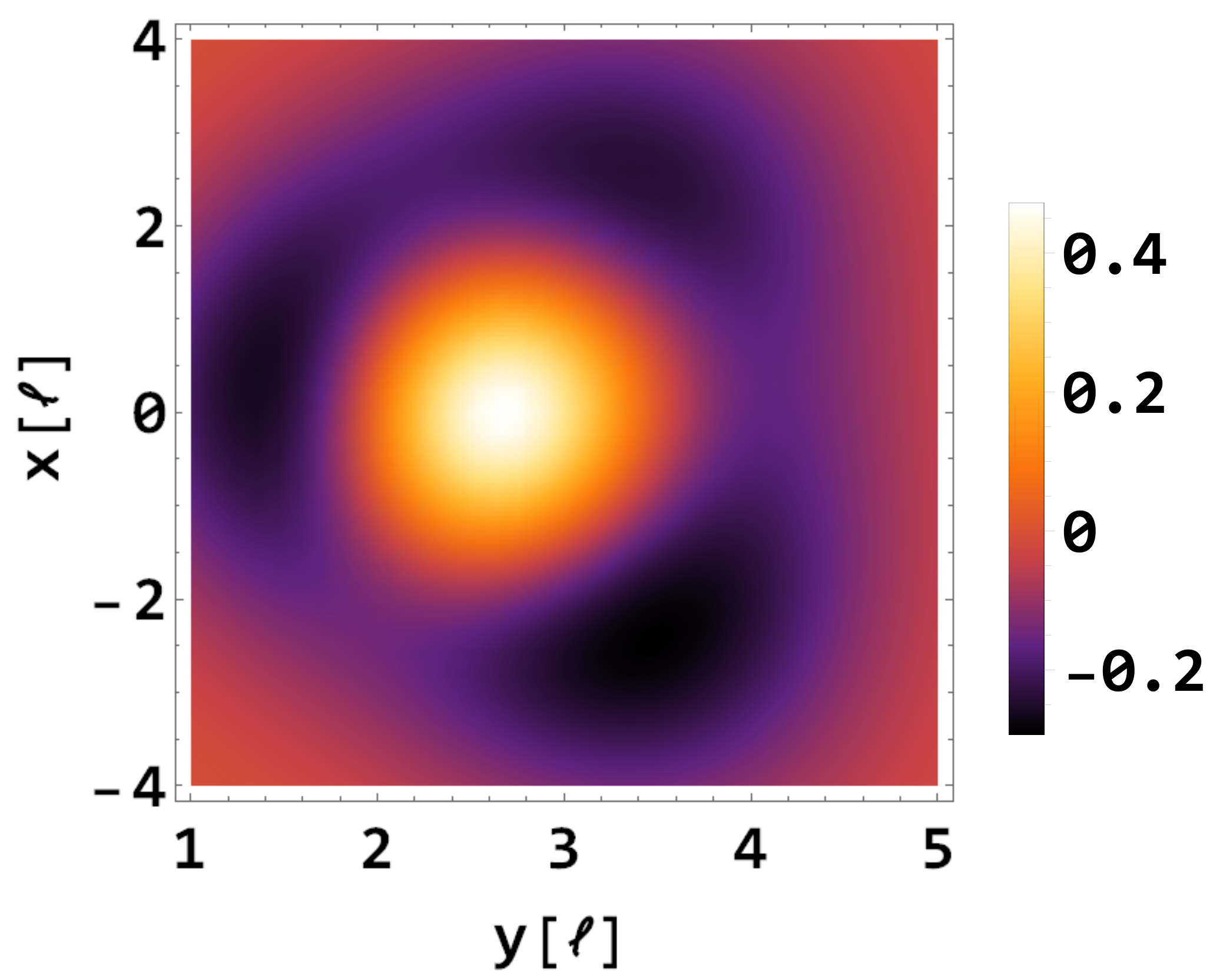} &
\includegraphics[width=.25\textwidth]{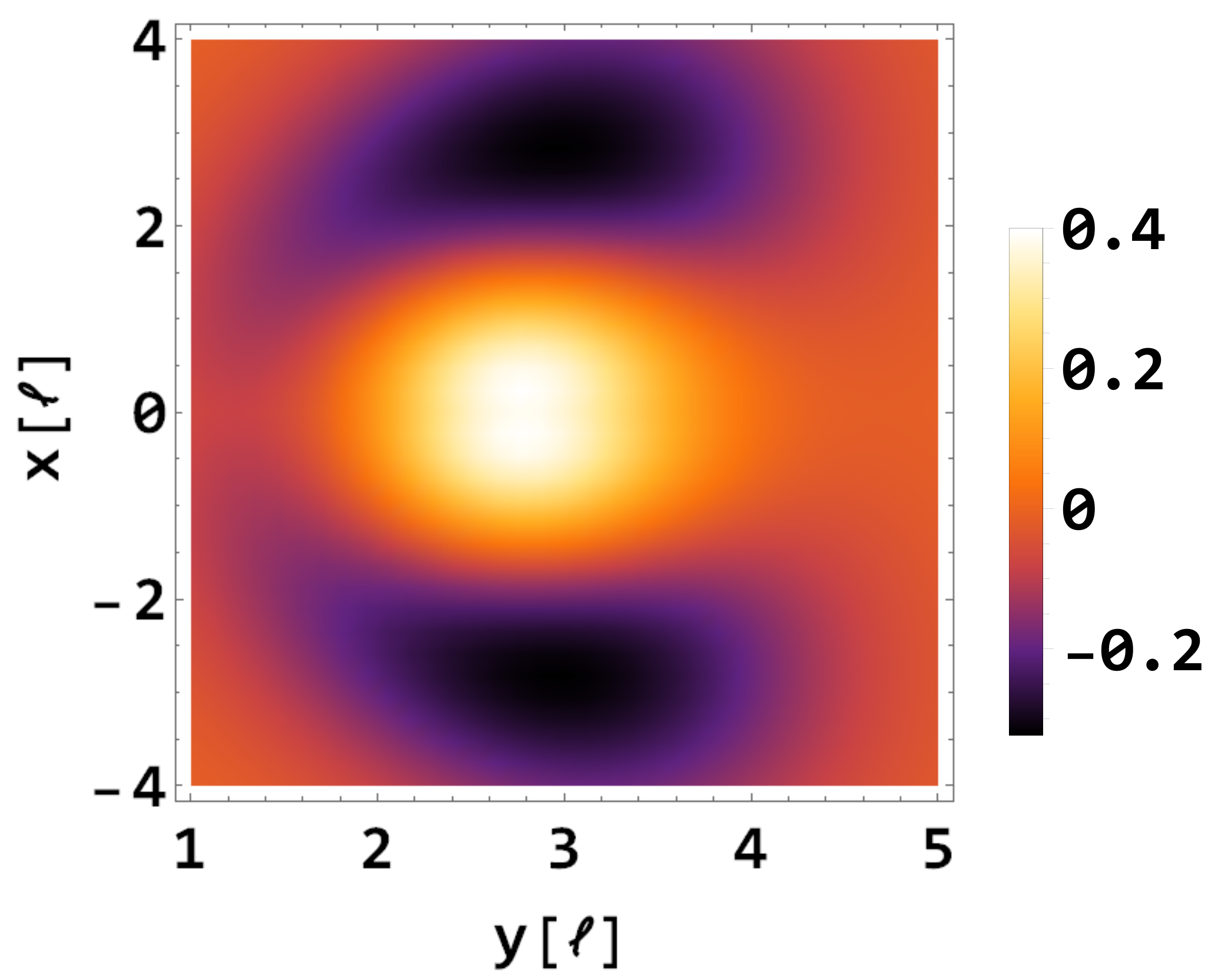} &
\includegraphics[width=.25\textwidth]{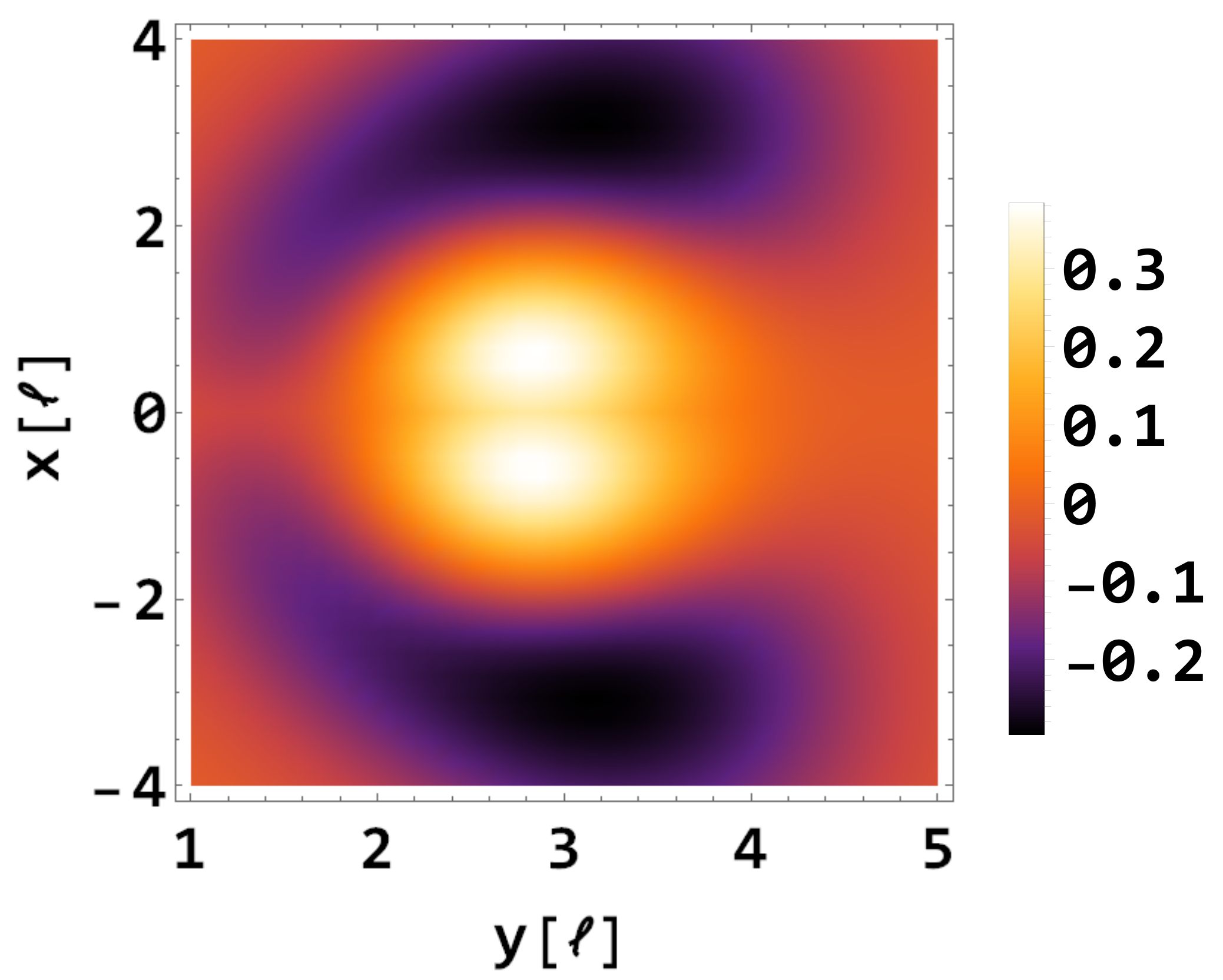} \\
\end{tabular}
\caption{Wave functions of the Hamiltonian \eqref{ham} with $p=0.795$ and a hard wall at $x=11\ell$ at various coupling strength, $g$ (in units of $\hbar\omega\ell$). Here $x=x_1-x_2$ is the relative motion coordinate and $y=\frac{x_1+x_2}{2}$ is the center-of-mass motion coordinate.}\label{wf}
\end{figure}

The first step to integrate Eq.\eqref{tdse} is to obtain its initial solution, i.e. $\psi(x_1,x_2,t=0)$. As a reference point, we consider the situation where the two-atom system is initially prepared in a deep trap with $p=0.795$. To compute eigenstates and eigenenergies in such a trap we put a hard wall, as in \cite{gharashi}, at $x=11\ell$ (Fig.\ref{trap_fig}) and assume that both atoms feel the same magnetic field gradient $\mathcal{C}=1894.18$ G/m. This set of parameters mimics the initial state preparation used in the experiments \cite{zuern,zuern2}. The corresponding stationary Sch\"{o}dinger equation is solved by means of the method from \cite{ishmukh3}. The energy spectrum in such a trap is shown in Fig.\ref{energy}. We focus on the excited state branches - upper and lower ones. These branches correspond to doubly excited states with respect to relative and center-of-mass motions \cite{ishmukh}. The nodal patterns of the wave functions is shown in Fig.\ref{wf}. At negative coupling ($g=-1$ in Fig.\ref{wf}) one clearly identifies the two nodes with respect to relative (upper branch) and center-of-mass (lower branch) motion coordinates. When we cross the point $g=0$, the nodal patterns of these states are mixed and at $g=1$ the doubly excited relative motion state, observed for the upper branch at $g=-1$, turns into the doubly excited center-of-mass motion state. The same situation, only vice versa, occurs for the lower branch: the doubly excited center-of-mass state turns into the doubly excited relative motion state. The origin of such nodal pattern mixing is due to rotational symmetry breaking at $g=0$: the anharmonic terms of the trap potential \eqref{trap} lift the doubly degenerate harmonic energy level \cite{ishmukh}. Similar effect has been observed in \cite{sala,ishmukh,noid}.

\onecolumngrid
\twocolumngrid

When computing the time-evolution of the wave function $\psi(x_1,x_2,t)$ at $t>0$, using \eqref{split}, we set $\mathcal{C}_1=1886$ G/m and $\mathcal{C}_2=1884$ G/m. These values approximately correspond to the values used for the tunneling dynamics used in \cite{gharashi} which models the actual values used in the experiments \cite{zuern,zuern2}. As was pointed out in \cite{gharashi} the accurate calculation of the parameters $\mathcal{C}$ should include Breit-Rabi formula. In the present paper we avoid such a consideration and expect that the qualitative picture of the whole tunneling analysis would not be altered significantly.

\subsection{Upper branch}

\begin{figure}[H]
\centering
\includegraphics[width=6cm,clip]{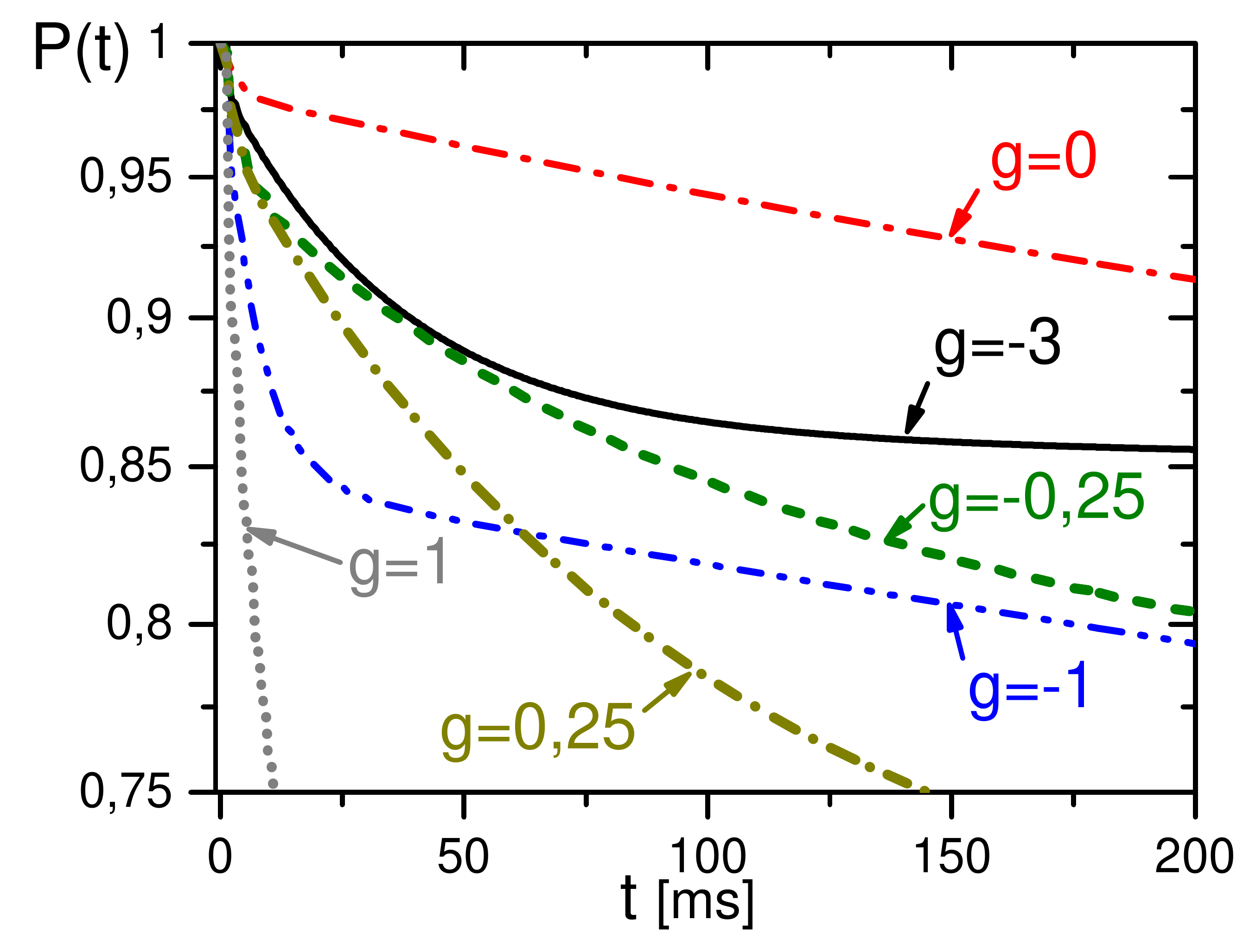}
\caption{(Color online) Total probability $P(t)$ for various coupling strengths $g$ (in units of $\hbar\omega\ell$)  with $p=0.73$.}\label{norm_upp}
\end{figure}

For an analysis of the tunneling dynamics we consider the total probability $P(t)$
\begin{eqnarray}\label{total}
P(t)=\iint\limits_{\Omega} dx_1 dx_2|\psi(x_1,x_2,t)|^2,
\end{eqnarray}
where, for the integration domain $\Omega$, we use the simulation grid $x_k\in[-3,40]$. Dependence of a decay rate of the total probability $P(t)$ goes in a non-monotonic manner as a function of the coupling strength $g$. From Fig.\ref{norm_upp}, one can notice that at $g=-0.25\hbar\omega$ the total probability $P(t)$ takes an intermediate value between the total probabilities $P(t)$ at $g=-3\hbar\omega$ and $g=-1\hbar\omega$. When we increase $g$ to $g=0$ the quantity $P(t)$ goes with an even smaller decay rate. Further increasing of $g$, however, leads to approximately monotonic dependence of the decay of $P(t)$ on $g$. This non-monotonic (and monotonic behavior discussed in the next section) behavior is somewhat similar to the behavior observed in \cite{ishmukh}.

To better understand the origin of such a non-monotonic behavior we pick for the initial wave functions the wave functions at $g=-1\hbar\omega\ell$, $g=0$ and $g=1\hbar\omega\ell$ and compute $P(t)$ for a wide range of the coupling strengths, starting from $g=-3\hbar\omega\ell$ up to $g=1\hbar\omega\ell$ (of course, the computed $P(t)$ differ significantly from $P(t)$, computed with the correct initial state wave functions, i.e. the initial states with the same coupling strengths $g$ as the value of $g$ used for $t>0$). We observe that, for the initial states with $g=0$ and $g=1\hbar\omega\ell$, the non-monotonic behavior still preserves: the decay rate of $P(t)$, starting from $g=-3\hbar\omega\ell$, decreases monotonically with increasing $g$, up to the decay rate of $P(t)$ at $g=0$, and then monotonically increases with increasing $g$. However, with these three different initial states we don't reproduce the non-monotonic dependence of $P(t)$ between the $g=-3\hbar\omega\ell$ and $g=-1\hbar\omega\ell$, i.e. $P(t)$ at $g=-0.25\hbar\omega\ell$ (Fig.\ref{norm_upp}). This suggests that the origin of such peculiar behavior, perhaps, refers to the initial wave function distribution.

\begin{figure}[H]
\centering
\includegraphics[width=6cm,clip]{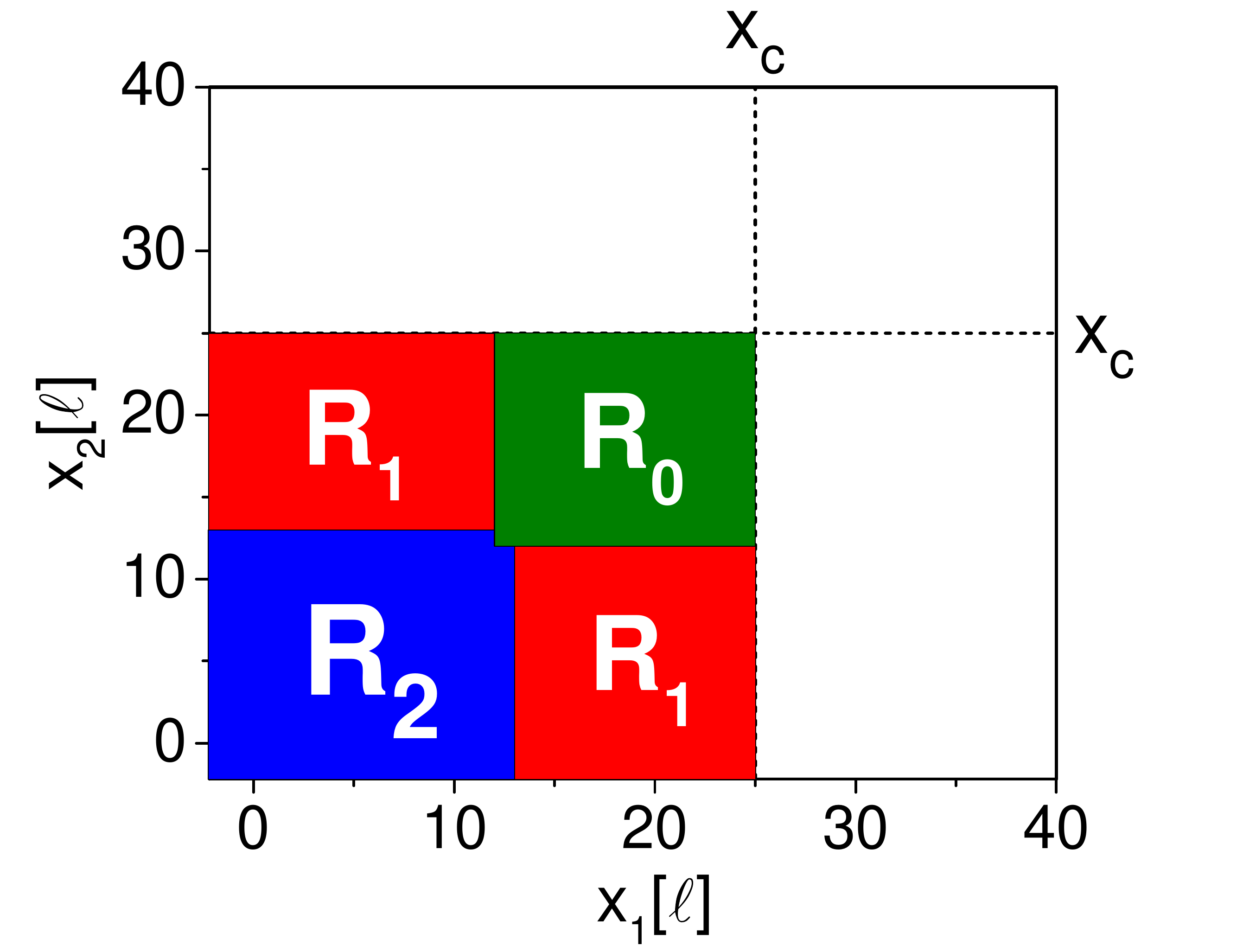}
\caption{(Color online) Partition of the configuration space into regions $R_2$ (the region that covers the trap size), $R_1$ and $R_0$. $x_c$ is the position at which CAP starts.}\label{regions}
\end{figure}

If we divide the whole configuration space into regions (Fig.\ref{regions}) and calculate the partial probabilities
\begin{eqnarray}\label{partial}
\nonumber
P_k(t)=\iint\limits_{region~k} dx_1 dx_2|\psi(x_1,x_2,t)|^2\\
\hspace{.3cm} k=0,1,2.
\end{eqnarray}
at each of the region then it is possible to extract the mean atom number that remains in the trap during the decay of the system. Region $R_2$ approximately covers the size of the two-atom trap (we take it as $x_j\in[-3,13]$) and therefore $P_2$ defines the probability to find the two atoms in the trap, $P_1$ defines the probability to find one atom in the trap and $P_0$ defines the probability that the two atoms escape from the trap. The mean atom number is defined as
\begin{eqnarray}\label{mean}
\bar{N}(t)=2P_2(t)+P_1(t).
\end{eqnarray}

\begin{figure}[H]
\centering
\includegraphics[width=6cm,clip]{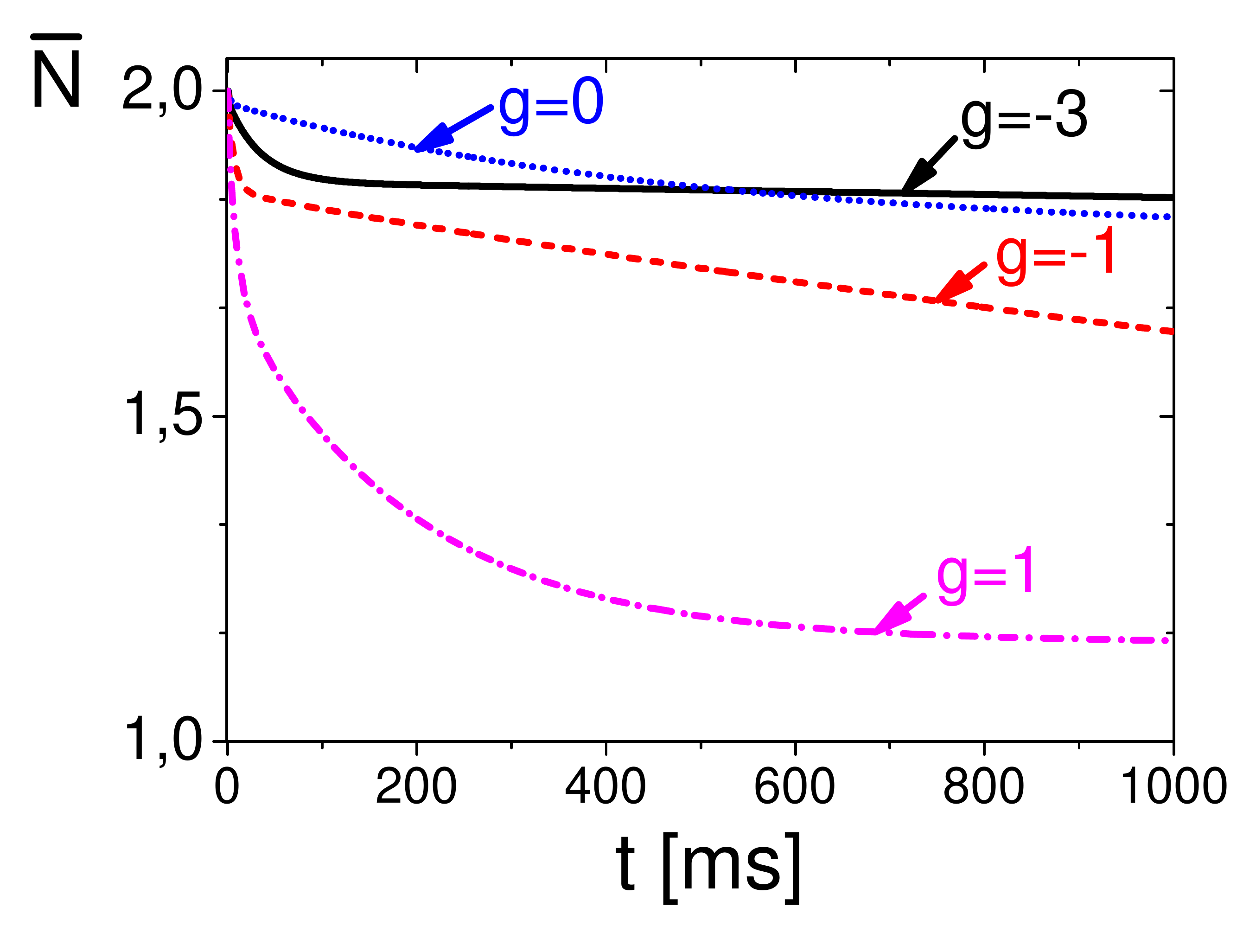}
\caption{(Color online) Mean particle number $\bar{N}(t)$ for various coupling strength $g$ (in units of $\hbar\omega\ell$).}\label{mean_upp}
\end{figure}

An analysis of the partial probabilities for the upper branch shows that the tunneling predominately goes into the regions $R_1$. Due to the normalization condition, $P_2(t)+P_1(t)+P_0(t)=1$, and that $P_0(t)$ is small, we approximate $P_1(t)$ as $P_1(t)\approx 1-P_2(t)$. Hence, the mean atom number is calculated as $\bar{N}(t)\approx P_2(t)+1$. The resulting $\bar{N}(t)$ is shown in Fig.\ref{mean_upp}.
Since the tunneling goes only into the regions $R_1$, within the considered time domain, we can conclude that only the first particle tunnels out of the trap (Fig.\ref{mean_upp}) while the second particle remains in the trap with very small decay rate \cite{gharashi}.

\subsection{Lower branch}
\begin{figure}[H]
\centering
\includegraphics[width=6cm,clip]{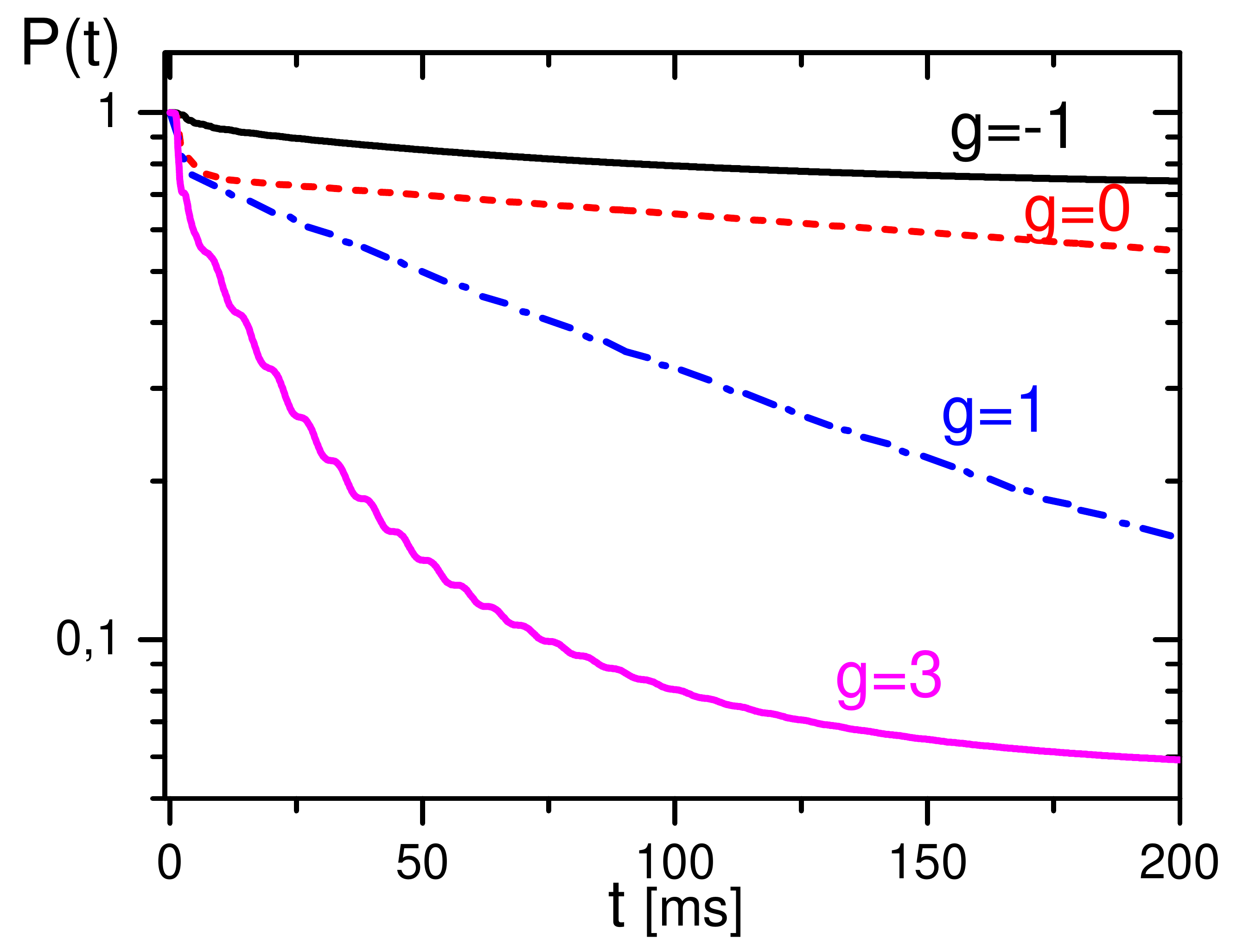}
\caption{(Color online) Total probability $P(t)$ for various coupling strengths $g$ (in units of $\hbar\omega\ell$) using $p=0.73$ at $t>0$}\label{norm_low}
\end{figure}

The decay rate from the lower branch of the excited states exhibits the monotonic dependence on the coupling strength $g$: the total probability $P(t)$ decays faster with increasing $g$ (Fig.\ref{norm_low}). From Fig.\ref{norm_low} one can observe a non-exponential decay of $P(t)$ at $g=3\hbar\omega\ell$. For high enough values of $g$, the decay from the lower branch goes faster than the decay from the upper branch of the excited states. This can be already noticed when comparing the decay of $P(t)$ at $g=0$ (cf. Fig.\ref{norm_upp}).

\begin{figure}[H]
\centering
\begin{tabular}{ccc}
\hspace{-.7cm}$g=-2\hbar\omega\ell$ &\hspace{-.7cm} $g=0$ &\hspace{-.7cm} $g=3\hbar\omega\ell$ \\
\multicolumn{3}{c}{$p=0.68$}\\
\hspace{-.34cm}\includegraphics[width=6cm,clip]{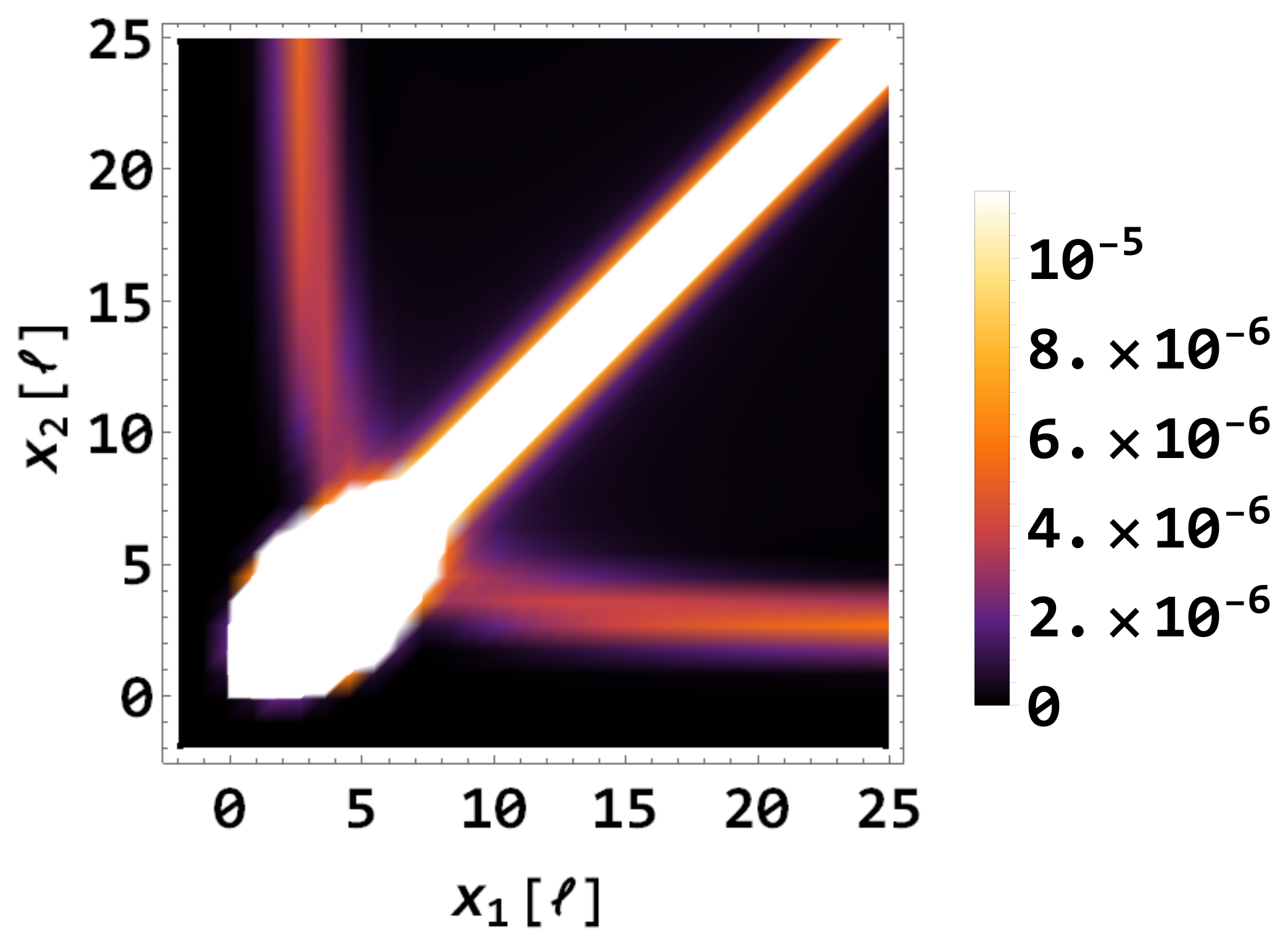} &
\hspace{-.3cm} \includegraphics[width=6cm,clip]{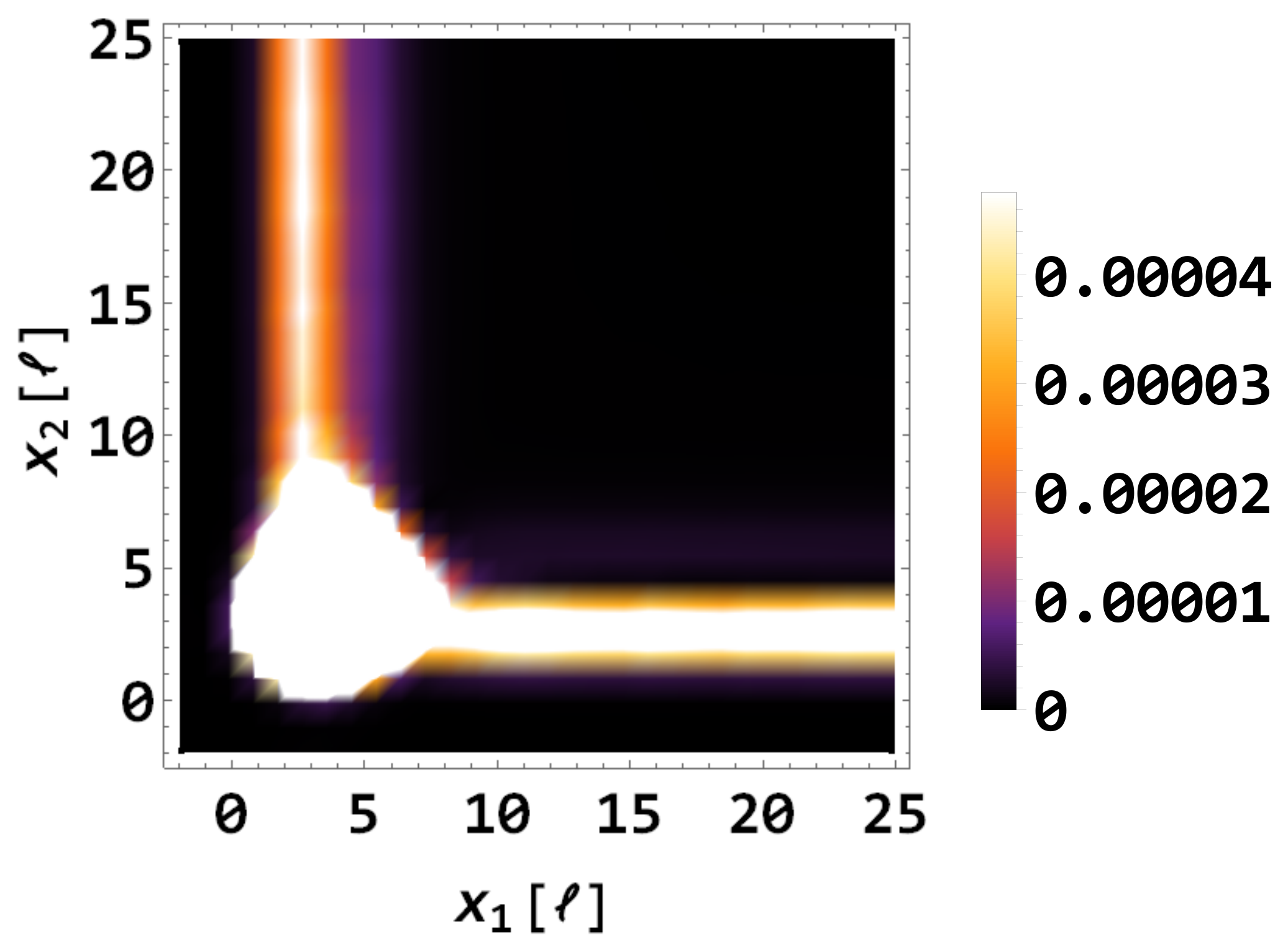} &
\hspace{-.3cm} \includegraphics[width=6cm,clip]{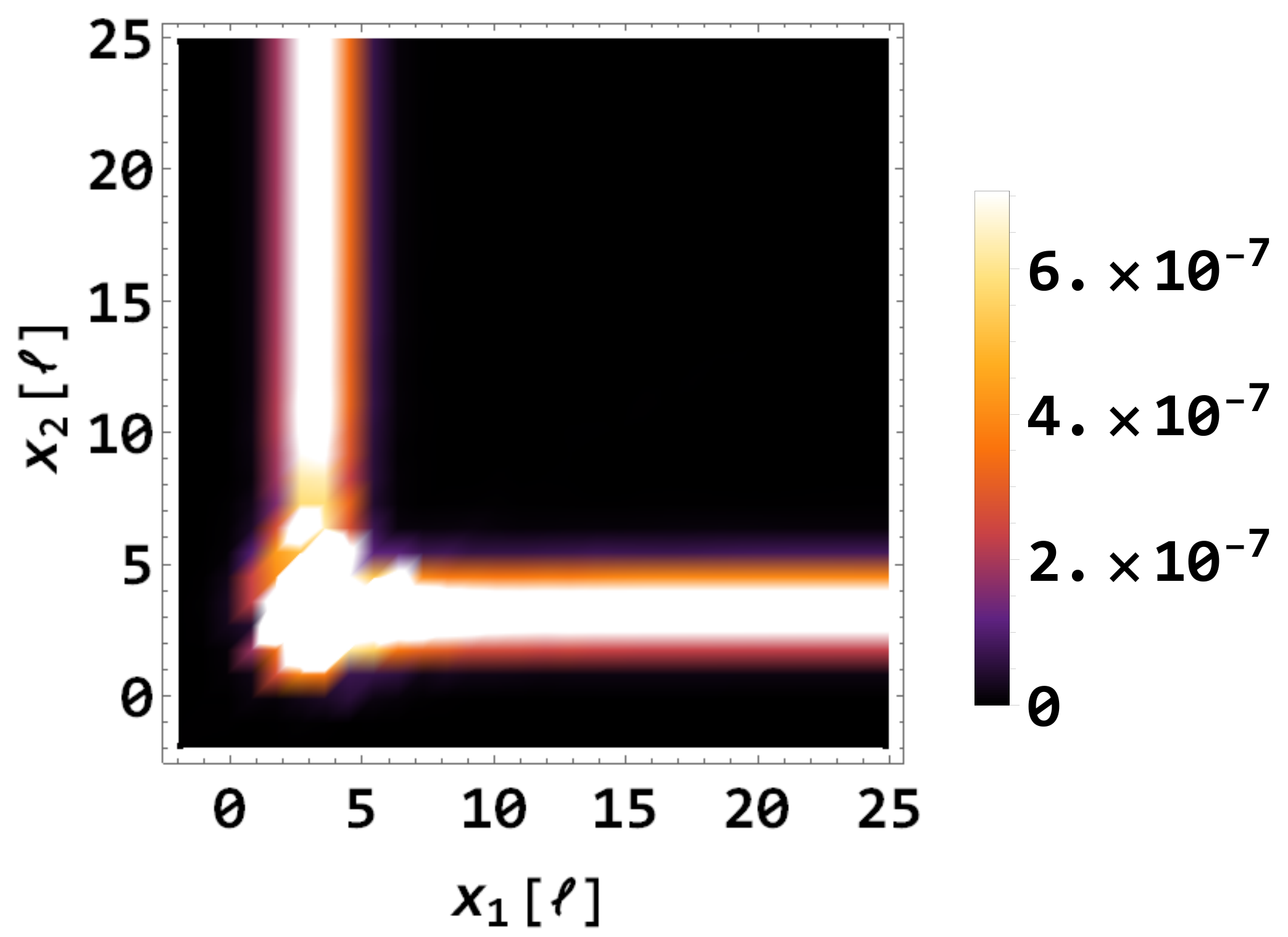} \\
\multicolumn{3}{c}{$p=0.73$}\\
\hspace{-.34cm} \includegraphics[width=6cm,clip]{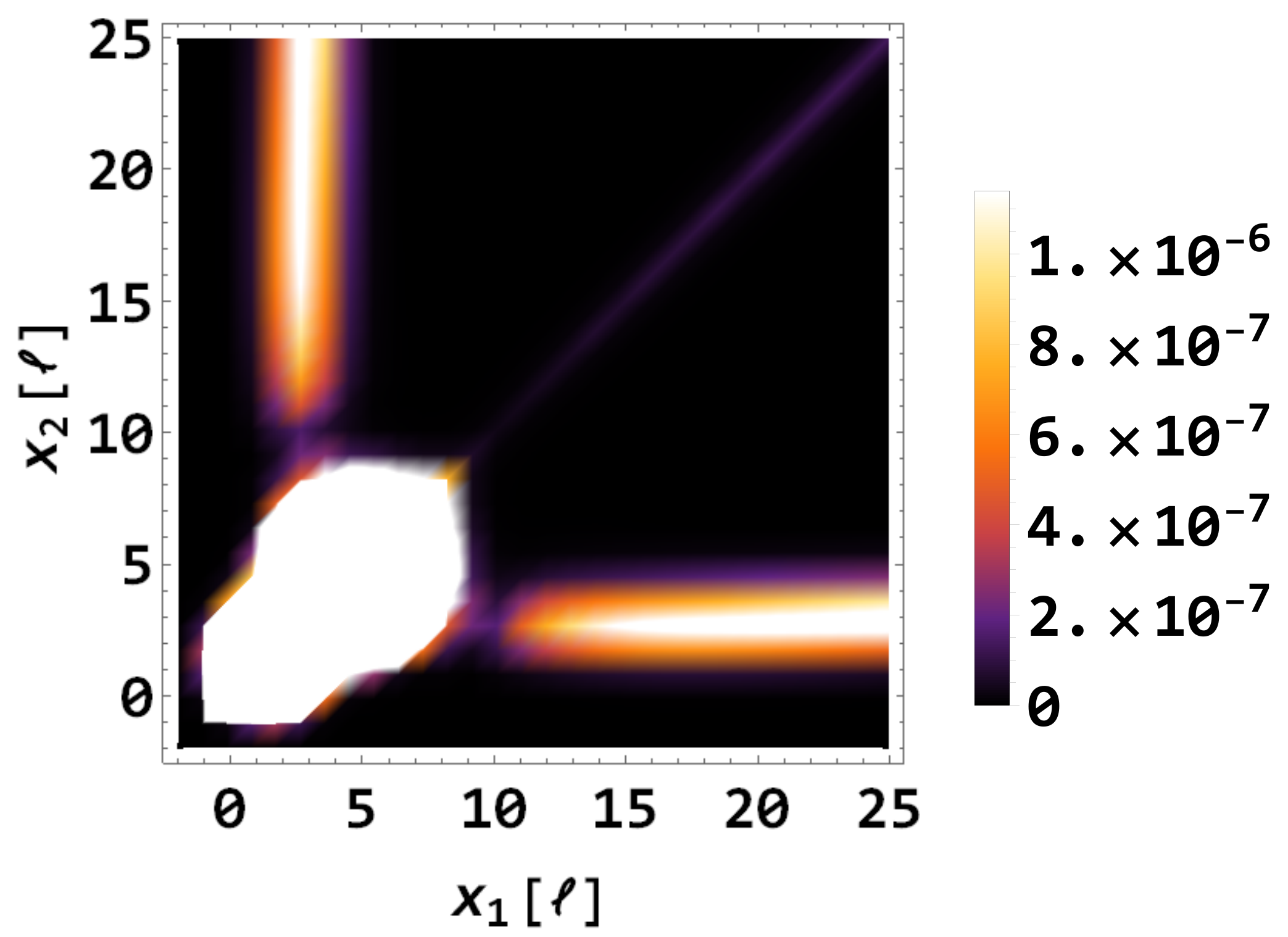} &
\hspace{-.3cm} \includegraphics[width=6cm,clip]{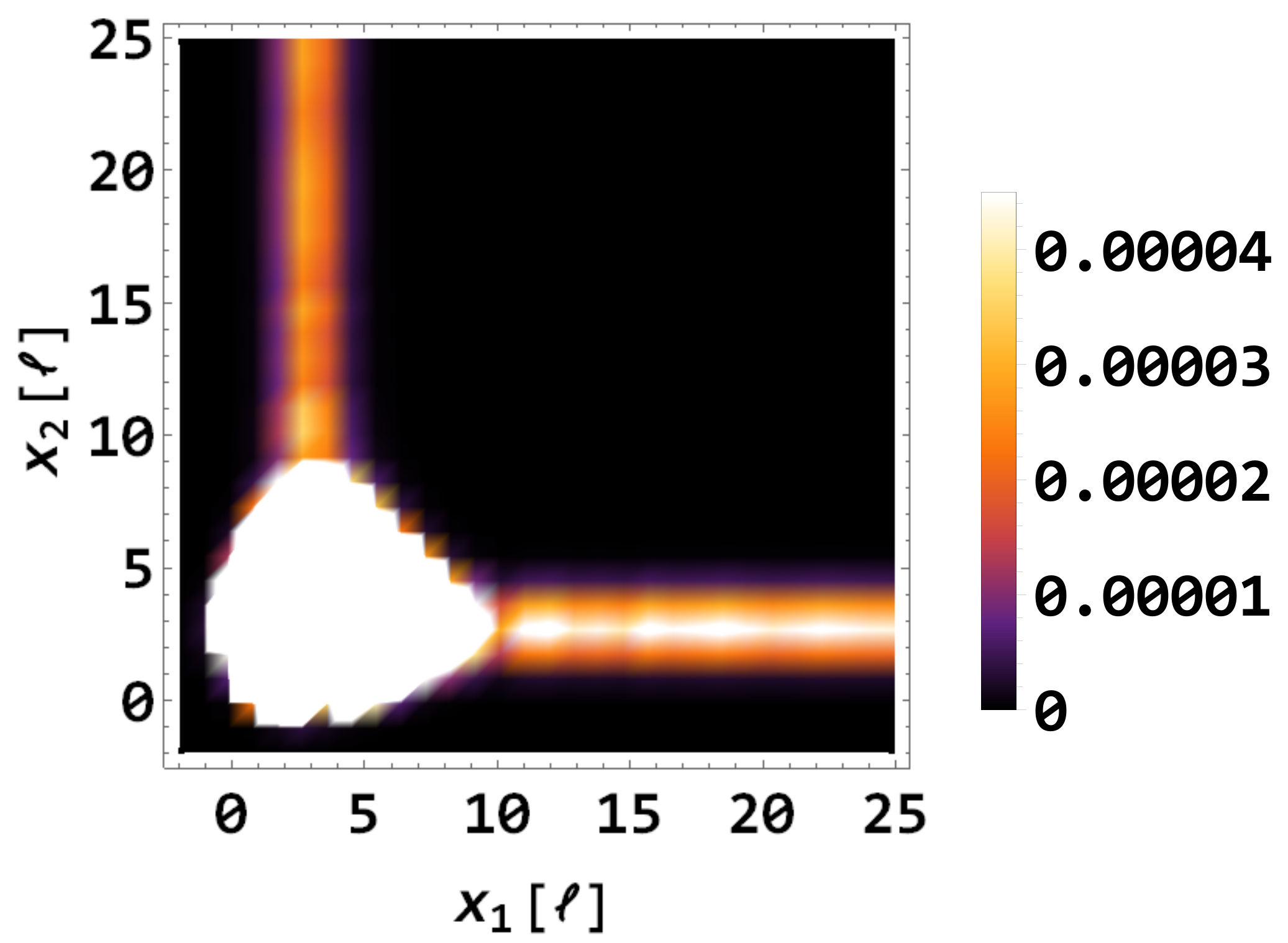} &
\hspace{-.3cm} \includegraphics[width=6cm,clip]{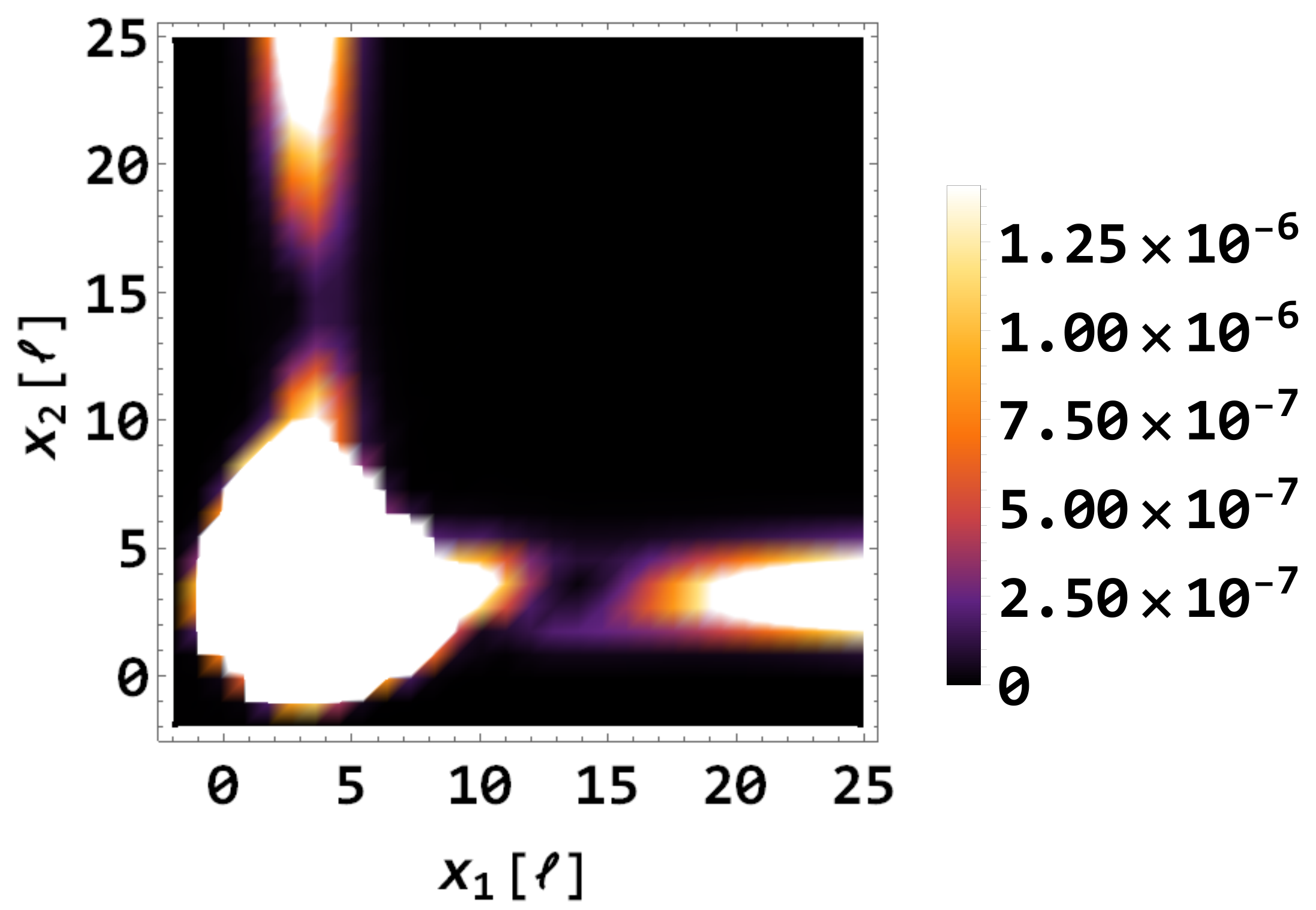} \\
\end{tabular}
\caption{(Color online) Probability current $|\textbf{j}(x_1,x_2,t)|$ of the lower excited state branch for the coupling strength values $g/(\hbar\omega\ell)=-2,0,3$ computed at $t=50$ ms. The values of the probability current shown in the legends are in $\omega/\ell$ units (note the different scale for all the graphs).}\label{flux_pic}
\end{figure}

Another interesting feature of the tunneling from this excited state branch occurs when we explore the transition from the correlated pair tunneling to uncorrelated one by decreasing the barrier, by using lower values of $p$ in \eqref{trap}. This transition takes place when the energy level of the excited bound state is close enough to the barrier level and there is strong enough attractive coupling $g$. We observe that the trap value $p=0.68$ (Eq.\eqref{trap}) is sufficient to observe such a transition. We identify the transition by computing the probability current
\begin{eqnarray}\label{flux}
\nonumber
j_k(x_1,x_2,t)=\frac{\hbar}{2mi}\left( \psi^{\ast}\frac{\partial \psi}{\partial x_k}-\psi\frac{\partial \psi^{\ast}}{\partial x_k} \right),\\
\hspace{.2cm}k=1,2.
\end{eqnarray}
From Fig.\ref{flux_pic}, with $p=0.68$, one can notice that $|\textbf{j}(x_1,x_2,t)|$ at $g=-2\hbar\omega\ell$ predominantly goes along the $x_1=x_2$ axis, which represents the tunneling of the two-atom system as a bound object, whereas at higher values of $g$, tunneling goes along the axes $x_1$ and $x_2$, which indicates a tunneling of the first particle. For higher values of $p$, $p=0.73$, this correlated tunneling channel is significantly suppressed and we observe that the sequential particle tunneling dominates also for the strong attraction, $g=-2\hbar\omega\ell$, and also repulsion. The tunneling rate of the second particle is expected to be negligible for the considered time domain \cite{gharashi}.

\onecolumngrid
\twocolumngrid

\section{CONCLUSION}
\label{conclusion}
We model the two-atom system with a realistic trap potential used in recent experiments \cite{zuern,zuern2}. We have extended the analysis made in \cite{gharashi} by considering tunneling from other excited states. It has been found that the decay rate from the upper excited state branch goes in a non-monotonic manner as a function of the coupling strength $g$, whereas the tunneling from the lower excited state branch exhibits monotonic dependence as a function of $g$.

We find how a transition from the correlated to uncorrelated tunneling behaves with changing the trap barrier. By decreasing the trap barrier, the correlated pair tunneling manifest itself at strong enough attractive coupling strength $g$. With increasing $g$ the two particles decay in a sequential manner. This transition is found to be absent for higher trap barrier. The particles from the upper branch, for the considered wide range of $g$, decay in the sequential way.

The analysis made in the paper can be extended to problems with more spatial degrees of freedom which can include transverse optical confinement \cite{gharashi}. It is also interesting to add more particles into the consideration.

\section*{ACKNOWLEDGMENTS}
The work was financially supported by the JINR Young Scientists and Specialists Grant (the grant number 18-302-04). We thank K. Hagino for helpful discussions.

\end{document}